\documentclass[twocolumn,tighten]{aastex63}

\usepackage{amsmath}
\usepackage{xspace}
\usepackage{multirow}
\usepackage{fancyapj}
\makeatletter
\def\restartappendixnumbering{\global\applettertrue
\setcounter{table}{0}
\setcounter{figure}{0}
\setcounter{equation}{0}
\def\thetable{\thesection\the\c@table}%
\renewcommand{\theHtable}{Supplement.\thetable}
\def\fnum@table{{\bf\tablename~\thetable}}%
\def\thefigure{\thesection\the\c@figure}%
\def\fnum@figure{{\bf\figurename~\thefigure}}%
}%
\makeatother

  {\list{}{\leftmargin=0.1in\rightmargin=0.1in}\item[]}%
  {\endlist}

\expandafter\def\csname editcolor1\endcsname{magenta}
\expandafter\def\csname editcolor2\endcsname{red}

\newcommand{\rbr}[1]{\ensuremath{\left( #1 \right)} }

\newcommand{\E}[1]{\ensuremath{\times 10^{#1}} }

\newcommand{\msol}{\ensuremath{M_{\odot}}\xspace}

\newcommand{\cts}{\rm\,ct\,\per{s}\xspace}
\newcommand{\kev}{\rm\,keV\xspace}
\newcommand{\hz}{\rm\,Hz\xspace}
\newcommand{\ks}{\rm\,ks\xspace}
\newcommand{\km}{\rm\,km\xspace}

\newcommand{\kpc}{\rm\,kpc\xspace}
\newcommand{\s}{\rm\,s\xspace}
\newcommand{\hr}{\rm\,hr\xspace}
\newcommand{\per}[1]{\rm\,#1\ensuremath{^{-1}}\xspace}
\newcommand{\persq}[1]{\rm\,#1\ensuremath{^{-2}}\xspace}

\newcommand{\fluxcgs}{{\rm\,erg{\per{s}}\persq{cm}\xspace}}

\newcommand{\no}{{\tt \#}}

\newcommand{\rxte}{\textrm{RXTE}\xspace}

\newcommand{\swift}{\textrm{Swift}\xspace}

\newcommand{\nicer}{\textrm{NICER}\xspace}
\newcommand{\integral}{\textrm{INTEGRAL}\xspace}
\newcommand{\astrosat}{\textrm{AstroSat}\xspace}
\newcommand{\src}{XTE~J1739\xspace}
\newcommand{\srcfull}{XTE~J1739--285\xspace}

\begin{document}
%\nolinenumbers

\title{The X-ray bursts of XTE J1739--285: a NICER sample}

\author{Peter Bult}
\affiliation{Department of Astronomy, University of Maryland,
  College Park, MD 20742, USA}
\affiliation{Astrophysics Science Division, 
  NASA's Goddard Space Flight Center, Greenbelt, MD 20771, USA}

\author{Diego Altamirano}
\affiliation{Physics \& Astronomy, University of Southampton, 
  Southampton, Hampshire SO17 1BJ, UK}

\author{Zaven Arzoumanian} 
\affiliation{Astrophysics Science Division, 
  NASA's Goddard Space Flight Center, Greenbelt, MD 20771, USA}

\author{Anna V. Bilous}
\affiliation{ASTRON, the Netherlands Institute for Radio Astronomy, 
  Postbus 2, 7990 AA Dwingeloo, The Netherlands}

\author{Deepto Chakrabarty}
\affil{MIT Kavli Institute for Astrophysics and Space Research, 
  Massachusetts Institute of Technology, Cambridge, MA 02139, USA}

\author{Keith C. Gendreau} 
\affiliation{Astrophysics Science Division, 
  NASA's Goddard Space Flight Center, Greenbelt, MD 20771, USA}

\author{Tolga G{\"u}ver}
\affiliation{Department of Astronomy and Space Sciences, Science Faculty, 
  Istanbul University, Beyaz{\i}t, 34119 Istanbul, Turkey}
\affiliation{Istanbul University Observatory Research and Application Center, 
  Beyaz{\i}t, 34119 Istanbul, Turkey}

\author{Gaurava K. Jaisawal}
\affil{National Space Institute, Technical University of Denmark, 
  Elektrovej 327-328, DK-2800 Lyngby, Denmark}

\author{Erik Kuulkers}
\affil{European Space Agency, 
  ESTEC, 2201 AZ Noordwijk, The Netherlands}

\author{C.~Malacaria}
\affiliation{NASA Marshall Space Flight Center, NSSTC, 320 Sparkman Drive, Huntsville, AL 35805, USA}\thanks{NASA Postdoctoral Fellow}
\affiliation{Universities Space Research Association, Science and Technology Institute, 320 Sparkman Drive, Huntsville, AL 35805, USA}

\author{Mason Ng}
\affil{MIT Kavli Institute for Astrophysics and Space Research, 
  Massachusetts Institute of Technology, Cambridge, MA 02139, USA}

\author{Andrea Sanna}
\affiliation{Dipartimento di Fisica, Universit\`a degli Studi di Cagliari, SP Monserrato-Sestu km 0.7, 09042 Monserrato, Italy}
\affiliation{INAF - Osservatorio Astronomico di Cagliari, via della Scienza 5, 09047 Selargius (CA), Italy}

\author{Tod E. Strohmayer} 
\affil{Astrophysics Science Division and Joint Space-Science Institute,
  NASA's Goddard Space Flight Center, Greenbelt, MD 20771, USA}

\begin{abstract}
  In this work we report on observations with the Neutron Star Interior
  Composition Explorer of the known neutron star X-ray transient \srcfull. We
  observed the source in 2020 February and March, finding it in a highly active
  bursting state. Across a 20-day period, we detected 32 thermonuclear X-ray
  bursts, with an average burst recurrence time of $2.0^{+0.4}_{-0.3}\hr$.
  A timing and spectral analysis of the ensemble of X-ray bursts reveals
  homogeneous burst properties, evidence for short-recurrence time bursts, and the
  detection of a 386.5\hz burst oscillation candidate.  The latter is especially notable,
  given that a previous study of
  this source claimed a 1122\hz burst oscillation candidate. We did not find any
  evidence of variability near 1122\hz, and instead find that the 386.5\hz
  oscillation is the more prominent signal of the two burst oscillation
  candidates. Hence, we conclude it is unlikely that \srcfull has
  a sub-millisecond rotation period. 
\end{abstract}


\section{Introduction}
  \label{sec:intro}
  %\linenumbers

  The neutron star low mass X-ray binary (LMXB) \srcfull (\src) was first
  discovered in 1999 October \citep{Markwardt1999} with the {Rossi X-ray
  Timing Explorer} (\rxte).
  It is perhaps best known for the 1122\hz burst oscillation candidate reported
  by \citet{Kaaret2007}. Since burst oscillations closely track the stellar
  spin frequency \citep[see, e.g.,][for a review]{Watts2012}, an 1122\hz
  oscillation would imply that \src is the fastest spinning neutron star
  currently known, and the only neutron star to spin at a sub-millisecond
  period. Other analyses of the same data, however, found no significant
  burst oscillation signals \citep{Galloway2008, Bilous2019}, suggesting that 
  the 1122\hz frequency reported by \citet{Kaaret2007} may have been a spurious
  result.

  Since its discovery, \src has shown an irregular pattern of X-ray outbursts.
  Galactic bulge scans performed with \rxte indicate that during its 1999
  outburst, the $2-10\kev$ source flux evolved between a minimum of about
  1\E{-9}\fluxcgs and a maximum of $5\E{-9}\fluxcgs$ over a period of roughly two
  weeks \citep{Markwardt1999}.  Additional weak outbursts occurred in 2001 and
  2003 \citep{Kaaret2007}, but only the former was followed-up with a limited
  number of pointed \rxte observations.
  No X-ray bursts were detected in these \rxte observations, however, leaving
  the nature of the source unknown.
  Only when the source once again became active in 2005 did it garner wider
  interest. In August 2005, the source was detected with \integral at a
  $3-10\kev$ flux of $\approx2\E{-9}\fluxcgs$ \citep{ATelBodaghee05}. About a
  month later, however, the flux had dropped to $2\E{-10}\fluxcgs$
  \citep{ATelShaw05}, and the first detection of thermonuclear (Type I) X-ray bursts
  demonstrated that the system harbors a neutron star \citep{ATelBrandt05}.
  Further observations with \rxte showed that the flux evolved between
  $4\E{-10}\fluxcgs$ and $1.5\E{-9}\fluxcgs$ across October and November, and
  after a period of Solar occultation, \src was found to still be visible in
  early 2006 \citep{ATelChenevez06}. Since then, outburst activity has been
  detected in 2012 \citep{ATelSanchez12}, and 2019 \citep{ATelMereminskiy19,
  ATelBult19e}, but has not been studied in detail.
  
  Over its long and rich history of outbursts, \src has shown a large number of
  Type I X-ray bursts, with 43 events cataloged in the Multi-Instrument Burst
  Archive (MINBAR, \citealt{Galloway2020}). Most these X-ray bursts we detected
  in \integral/JEM-X, and only six of them have been observed with \rxte.
  Hence, the sample of X-ray bursts with sufficient time-resolution to allow
  for a burst oscillation search is limited, and has not changed since the
  original analysis of \citet{Kaaret2007}.

  On 2020 February 8, \integral detected a brightening of \src
  \citep{ATelSanchez20}, which was quickly confirmed to be entering a new
  outburst cycle with a follow-up \swift/XRT observation \citep{ATelBozzo20}.
  In an effort to increase the sample of high-fidelity X-ray bursts from
  this target, we commenced regular monitoring with the {Neutron Star
  Interior Composition Explorer} (\nicer, \citealt{Gendreau2017}) on 2020
  February 13. This effort has yielded an extensive dataset on this source, including
  the detection of 32 X-ray bursts. In this paper we present the spectral and
  timing analysis of these events.

\section{Observations}
\label{sec:observations}
  We observed \src with \nicer between 2020 February 13 and 2020 April 4 for an
  integrated unfiltered exposure of {335}\ks.  These data are available
  under ObsIDs $20502801nn$ and $30502801mm$, where $nn$ runs from 25 through 36
  and $mm$ from 01 through 30. All data were processed and calibrated using
  \textsc{nicerdas} version 7a, which is available as part of \textsc{heasoft}
  version 6.27.2. We screened the data using standard cleaning criteria,
  retaining only those time intervals during which the pointing offset was
  $<54\arcsec$, the elevation with respect to bright Earth limb was
  $>30\arcdeg$, the angle relative to the dark Earth limb was $>15\arcdeg$, and
  the instrument was outside the South Atlantic Anomaly (SAA).  By default, the
  \textsc{nicerdas} processing further screens epochs of increased background
  by filtering on the rate of saturating particles (overshoots). However, this
  method was found to be affected by statistical fluctuations in overshoot
  rate, leading to spurious 1-second gaps in the light curve. Following
  \citet{Bult2020a}, we corrected for this effect by applying a 5-second
  smoothing average to the overshoot rate before evaluating the screening criteria.
  After applying these screening filters we were left with {244\ks} of good
  time exposure.

  Inspecting a light curve of the clean data, we readily identified {31}
  X-ray bursts.  
  Additionally, we observed {one} weak burst-like feature which we will
  call a mini-burst. This event is investigated separately in Section
  \ref{sec:recurrence}.
  One of the X-ray bursts (\no29) was found to suffer from a telemetry issue,
  causing a large number of sub-second gaps in the raw data.  Due to \nicer's
  modular design, however, the different detectors showed gaps at different times.
  We processed the seven Measurement/Power Units (MPUs) separately, and computed
  a light curve for each of them. By summing these light curves, weighted by
  their effective exposure in each time-bin, we recovered an uninterrupted X-ray
  burst light curve.  
  Nonetheless, due to the uneven sampling, these data are unsuited for 
  timing analyses and are not included in Section \ref{sec:timing}.  Finally, we 
  inspected a light curve of the unfiltered
  data, and found that one additional X-ray burst occurred during SAA passage
  (\no31). To recover this particular burst, we reprocessed the relevant ObsID
  using a manually adjusted good-time interval table.  In Table
  \ref{tab:bursts} we list all 32 observed X-ray bursts and indicate which
  were affected by special observing conditions.

\section{Results}
\label{sec:results}

\subsection{Light curves}
\label{sec:light curves}
\begin{figure}[t]
  \centering
  \includegraphics[width=\linewidth]{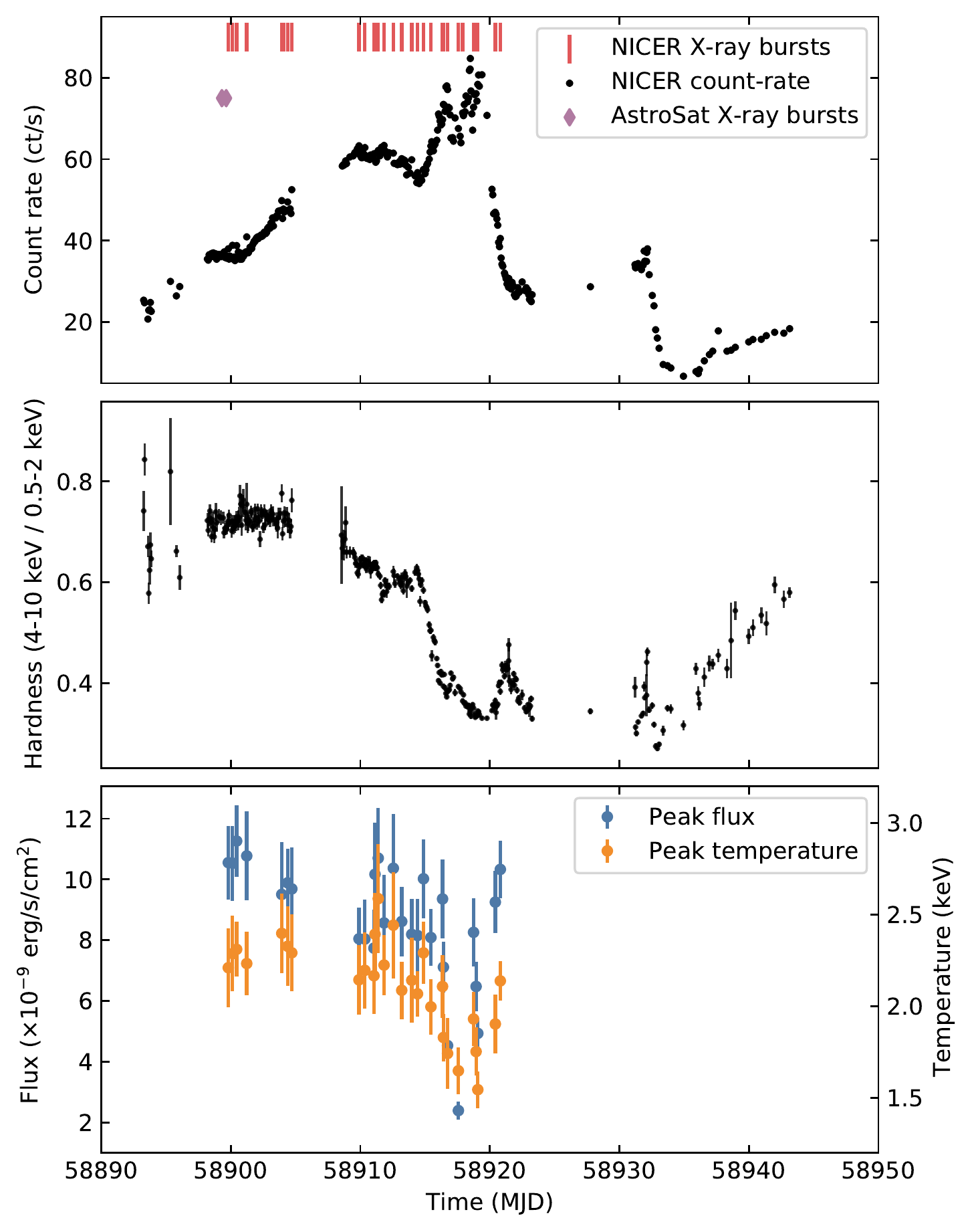}
  \caption{%
	Top: \srcfull light curve in the $0.5-10\kev$ energy range, showing one point
	per \nicer pointing. X-ray bursts have been filtered out of the light curve and are
	indicated with red bars instead.  The purple diamonds indicate the times
	of two X-ray bursts observed with \astrosat \citep{ATelChakraborty20}. 
	Middle: hardness ratio calculated as the $4-10$\kev rate over the $0.5-2$\kev rate.
	Bottom: peak bolometric flux (left) and peak blackbody temperature (right)
	for each of the \nicer X-ray bursts. Error bars show the 90\% confidence region.
  }
  \label{fig:light curve}
\end{figure}

  Over the course of our two-month monitoring campaign, \src showed large
  swings in its mean $0.5-10\kev$ count-rate. During the first $\approx30$ days
  of monitoring, the mean rate gradually increased from $20\cts$ to a peak of
  $80\cts$. On 2020 March 11 the observed rate showed a sudden drop, decreasing
  by more than half, to approximately $30\cts$. Over the following 2 weeks, the
  source showed a modest intensity increase, before plummeting again on 2020
  March 23. Beyond this date the source intensity steadily increased again, but
  visibility and scheduling constraints prevented further high cadence
  monitoring.  We show the complete light curve in the top panel of Figure
  \ref{fig:light curve}, where we removed the X-ray bursts to highlight the
  evolution of the mean source rate. In this figure, we also show the evolution
  of the hardness ratio, which we calculated as the $4-10\kev$ energy band count-rate
  divided by the $0.5-2\kev$ band rate. 
  We find that for the first 25 days of observations this hardness ratio 
  evolves roughly in anti-correlation with the $0.5-10\kev$ count-rate. Around
  the time the source count-rate peaks, the hardness ratio transitions from a high
  to a low value and becomes positively correlated with the source count-rate.
  
  A total of 32 X-ray bursts were observed over the course of our monitoring
  campaign, all of which occurred during a 20-day window in which the mean rate was
  above $\approx35\cts$ (and the X-ray flux was $\geq4.5\E{-10}\fluxcgs$; see
  Section \ref{sec:spectroscopy}). We indicated the times of these bursts in Figure
  \ref{fig:light curve} using vertical red bars.
  Five of the 32 observed bursts were only observed partially:
  bursts \no6, \no17, and \no31 were truncated to various degrees, whereas for
  bursts \no13 and \no26 we missed the onset. 

  We determined the start time of each burst using a two-step procedure. 
  First, we constructed a 1-s time resolution light curve for every \nicer
  pointing that contained an X-ray burst. We then searched this light curve for
  the first bin whose count-rate exceeded the averaged count-rate by a factor of
  2. 
  The averaged count-rate was calculated over a 20-s long window, which we
  separated from the test bin by a $10\s$ shift. That is, when testing bin
  $t_n$, we calculated the light curve average over $[t_{n-30}, t_{n-10}]$.
  If an insufficient number of bins was available before the X-ray
  bursts, we sampled the end of the light curve instead.
  This procedure correctly identified the burst onset in all cases, with the
  exception of bursts \no13 and \no26, for which the rise was not observed. The
  resulting burst start times, however, had a $\approx1\s$ uncertainty and
  failed to align all bursts. 
  In a second step, we therefore refined the onset times to improve the burst
  alignment.  For each burst, we constructed a light curve of the burst rise,
  which we interpolated using a first-order Savitzky-Golay filter over
  a 3-s window \citep{SavGol}. We then subtracted the mean persistent
  count-rate and determined the time at which the burst count-rate passed
  through $50\cts$, which is about 10\% of the mean peak burst
  count-rate. We defined this intersection as the onset time of the burst. All
  resulting onset times are listed in Table
  \ref{tab:bursts}.
  
\begin{table*}[t]
    \caption{%
        X-ray burst overview
        \label{tab:bursts}
    }
	\hspace*{-3.2cm}
	\newcommand{\mb}[1]{\phantom{\tablenotemark{b}}#1\tablenotemark{b}}
	\newcommand{\mc}[1]{\phantom{\tablenotemark{c}}#1\tablenotemark{c}}
	\newcommand{\md}[1]{\phantom{\tablenotemark{d}}#1\tablenotemark{d}}
    \begin{tabular}{c c c c c c c c c c c}
	    Burst & ObsID\tablenotemark{a} & Onset Time & Onset Date & Peak Flux & Fluence & $\Delta t_\text{rec}$ & $\Delta t_\text{rise}$ & $\epsilon$ & $\tau$ & $\alpha$ \\
	    ~     & ~                  & (MET)      & (MJD)  & ({\E{-8}}erg/s/cm$^2$) & ({\E{-7}}erg/cm$^2$) & (hours) & (s) & (s)  & (s) & ~    \\
        \tableline
	~1     & x30 & 193692577   & 58899.81223 & $1.06 \pm 0.12$ & $2.10 \pm 0.04$ &   3.88 &  5.2 & 19.0 & $20$ &   50  \\  % \pm 2
	~2     & x31 & 193720395   & 58900.13418 & $1.05 \pm 0.12$ & $2.24 \pm 0.05$ &   7.73 &  3.9 & 18.5 & $21$ &   99  \\  % \pm 3
	~3     & x31 & 193748118   & 58900.45505 & $1.13 \pm 0.12$ & $2.22 \pm 0.04$ &   7.70 &  4.8 & 15.1 & $20$ &   97  \\  % \pm 2
	~4     & x32 & 193815367   & 58901.23340 & $1.08 \pm 0.15$ & $1.87 \pm 0.04$ &  18.68 &  4.8 & 17.7 & $17$ &  279  \\  % \pm 2
	~5     & x34 & 194049124   & 58903.93892 & $0.95 \pm 0.17$ & $2.19 \pm 0.05$ &    -   &  7.0 & 22.2 & $23$ &    -  \\  % \pm 4 
	\mb{~6}& x35 & 194060568   & 58904.07138 & $1.43 \pm 0.22$ & $1.76 \pm 0.06$ &   3.18 &  8.3 &   -  &   -  &    -  \\  %  -    
	~7     & x35 & 194088338   & 58904.39278 & $0.99 \pm 0.11$ & $2.14 \pm 0.04$ &   7.71 &  7.9 & 19.6 & $22$ &  133  \\  % \pm 2
	~8     & x35 & 194115999   & 58904.71294 & $0.97 \pm 0.14$ & $2.09 \pm 0.04$ &   7.68 &  4.5 & 23.9 & $22$ &  142  \\  % \pm 3
	~9     & y01 & 194564001   & 58909.89815 & $0.80 \pm 0.10$ & $1.94 \pm 0.04$ &    -   &  6.2 & 22.9 & $24$ &    -  \\  % \pm 3 
	10     & y02 & 194603602   & 58910.35649 & $0.80 \pm 0.13$ & $1.94 \pm 0.04$ &  11.00 &  9.8 & 22.2 & $24$ &  257  \\  % \pm 4
	11     & y03 & 194663960   & 58911.05508 & $0.77 \pm 0.13$ & $1.97 \pm 0.04$ &  16.77 &  5.2 & 23.1 & $25$ &  368  \\  % \pm 4
	12     & y03 & 194670876   & 58911.13513 & $1.02 \pm 0.17$ & $2.00 \pm 0.04$ &   1.92 &  5.4 & 21.8 & $20$ &   43  \\  % \pm 3
	\mb{13}& y03 & 194686325   & 58911.31394 & $0.22 \pm 0.03$ & $0.63 \pm 0.01$ &   4.29 &   -  &  -   &   -  &    -  \\  %  -    
	14     & y03 & 194692869   & 58911.38968 & $1.07 \pm 0.16$ & $2.05 \pm 0.05$ &   1.82 &  3.6 & 21.1 & $19$ &   39  \\  % \pm 3
	15     & y03 & 194731453   & 58911.83625 & $0.86 \pm 0.16$ & $1.91 \pm 0.05$ &  10.72 &  5.3 & 21.9 & $22$ &  255  \\  % \pm 4
	16     & y04 & 194793276   & 58912.55180 & $1.04 \pm 0.18$ & $2.05 \pm 0.05$ &  17.17 &  6.3 & 18.4 & $20$ &  363  \\  % \pm 3
	\mb{17}& y05 & 194849315   & 58913.20039 & $0.86 \pm 0.11$ & $1.84 \pm 0.04$ &  15.57 &  6.9 &  -   &   -  &    -  \\  %  -    
	18     & y05 & 194916105   & 58913.97343 & $0.82 \pm 0.12$ & $1.89 \pm 0.04$ &  18.55 &  7.3 & 19.9 & $23$ &  405  \\  % \pm 3
	19     & y06 & 194954634   & 58914.41937 & $0.82 \pm 0.12$ & $2.01 \pm 0.04$ &  10.70 &  7.2 & 21.0 & $25$ &  215  \\  % \pm 4
	20     & y06 & 194994326   & 58914.87876 & $1.00 \pm 0.13$ & $2.07 \pm 0.04$ &  11.03 &  4.4 & 20.0 & $21$ &  215  \\  % \pm 3
	21     & y07 & 195044088   & 58915.45471 & $0.81 \pm 0.09$ & $1.87 \pm 0.03$ &  13.82 &  4.5 & 20.2 & $23$ &  327  \\  % \pm 3
	22     & y08 & 195121583   & 58916.35165 & $0.94 \pm 0.13$ & $1.66 \pm 0.04$ &  21.53 &  6.1 & 14.1 & $18$ &  691  \\  % \pm 2
	23     & y08 & 195127479   & 58916.41988 & $0.71 \pm 0.08$ & $1.45 \pm 0.03$ &   1.64 &  6.0 & 12.7 & $20$ &   65  \\  % \pm 2
	24     & y08 & 195155076   & 58916.73930 & $0.45 \pm 0.06$ & $1.15 \pm 0.03$ &   7.67 & 11.5 & 12.0 & $25$ &  409  \\  % \pm 3
	25     & y09 & 195227270   & 58917.57488 & $0.24 \pm 0.03$ & $0.44 \pm 0.01$ &  20.05 &  3.6 &  8.2 & $19$ & 2467  \\  % \pm 2
	\mb{26}& y09 & 195255276   & 58917.89902 & $0.07 \pm 0.01$ & $0.23 \pm 0.02$ &   7.78 &   -  &  -   &   -  &    -  \\  %  -    
	27     & y10 & 195328389   & 58918.74523 & $0.83 \pm 0.11$ & $1.56 \pm 0.03$ &  20.31 &  7.5 & 12.0 & $19$ &  776  \\  % \pm 3
	28     & y10 & 195345351   & 58918.94156 & $0.65 \pm 0.08$ & $1.19 \pm 0.02$ &   4.71 &  6.6 & 11.7 & $18$ &  254  \\  % \pm 2
	\mc{29}& y11 & 195350027   & 58918.99567 & $0.64 \pm 0.07$ & $1.32 \pm 0.04$ &   1.65 &  5.6 & 10.6 & $20$ &   80  \\  % \pm 2
	30     & y11 & 195356877   & 58919.07495 & $0.49 \pm 0.04$ & $0.76 \pm 0.02$ &   1.55 &  5.0 &  8.1 & $15$ &  134  \\  % \pm 1
	\md{31}& y12 & 195472719   & 58920.41571 & $0.93 \pm 0.10$ & $1.37 \pm 0.03$ &    -   &  5.4 &  7.1 & $15$ &    -  \\  % \pm 2 
	32     & y12 & 195506990   & 58920.81236 & $1.03 \pm 0.09$ & $1.78 \pm 0.03$ &   9.22 &  4.3 & 10.3 & $17$ &  159  \\  % \pm 2
        \tableline
    \end{tabular}
	\tablecomments{%
	  All reported flux measurements are unabsorbed. Burst onset times are reported
	  in \nicer's Mission Elapsed Time (MET).
	  The final five columns, $\Delta t_\text{rec}$, $\Delta t_\text{rise}$, $\epsilon$, $\tau$, and $\alpha$, give, respectively:
	  the wait time since the previous burst,
	  the rise time of the burst,
	  the decay e-folding time, 
	  the ratio of the burst fluence to the peak flux, and
	  the ratio of the integrated persistent flux to the burst fluence
	  (see text for definitions).
	  We only list the recurrence time if it is shorter than 1 day. 
	  The recurrence time of the first burst is calculated relative to an 
	  X-ray burst observed with \astrosat \citep{ATelChakraborty20}. 
	  Uncertainties are quoted at 90\% confidence. \\
	}
	\tablenotetext{a}{We only list the last two digits, so x=$20502801$ and y=$30502801$}
	\tablenotetext{b}{Partial burst.}
	\tablenotetext{c}{Fragmented.}
	\tablenotetext{d}{During SAA passage.}
\end{table*}

\begin{figure*}[t]
  \centering
  \includegraphics[width=\linewidth]{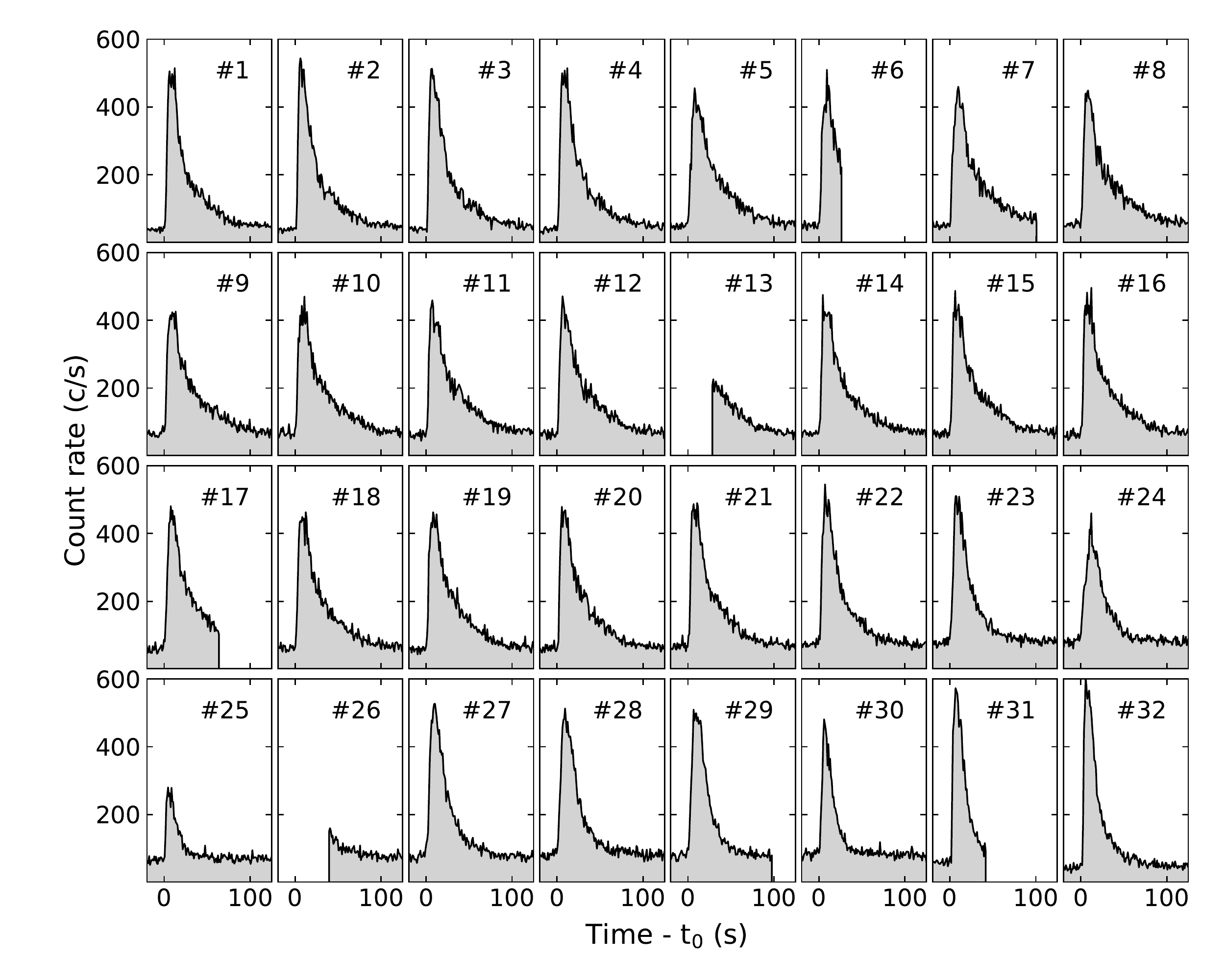}
  \caption{%
	Light curves of each individual X-ray burst from \srcfull
	observed with \nicer. These light curves are in the $0.5-10\kev$
	energy band, binned at 1-s time resolution, and expressed relative to
	$t_0$, the start time of each respective X-ray burst (see Table
	\ref{tab:bursts}). 
  }
  \label{fig:burst profiles}
\end{figure*}

  In Figure \ref{fig:burst profiles} we show the $0.5-10\kev$ 1-s light curves of all X-ray
  bursts relative to their respective onset. The light curve profiles are
  very similar across all X-ray bursts: the bursts take $5-7\s$ to rise to a peak rate
  of about $500\cts$ and have duration of approximately $100\s$. 
  To quantify the shapes of these profiles more precisely, 
  we measured the burst rise time as the interval between the burst onset and
  the first 1-s light curve bin that was within one standard deviation of the respective
  burst peak count-rate. We also determined the end time of each X-ray
  burst by finding the time at which the count-rate decayed back to the
  preburst level. Specifically, we scanned the 1-s time resolution light curve,
  starting from $t_0 + 10\s$, for the first time bin whose count-rate was
  within $1\sigma$ of the preburst rate. Considering all bursts that decayed
  before the end of their respective observations, we found an average burst
  duration of $95\pm14\s$. 

\begin{figure}[t]
  \centering
  \includegraphics[width=\linewidth]{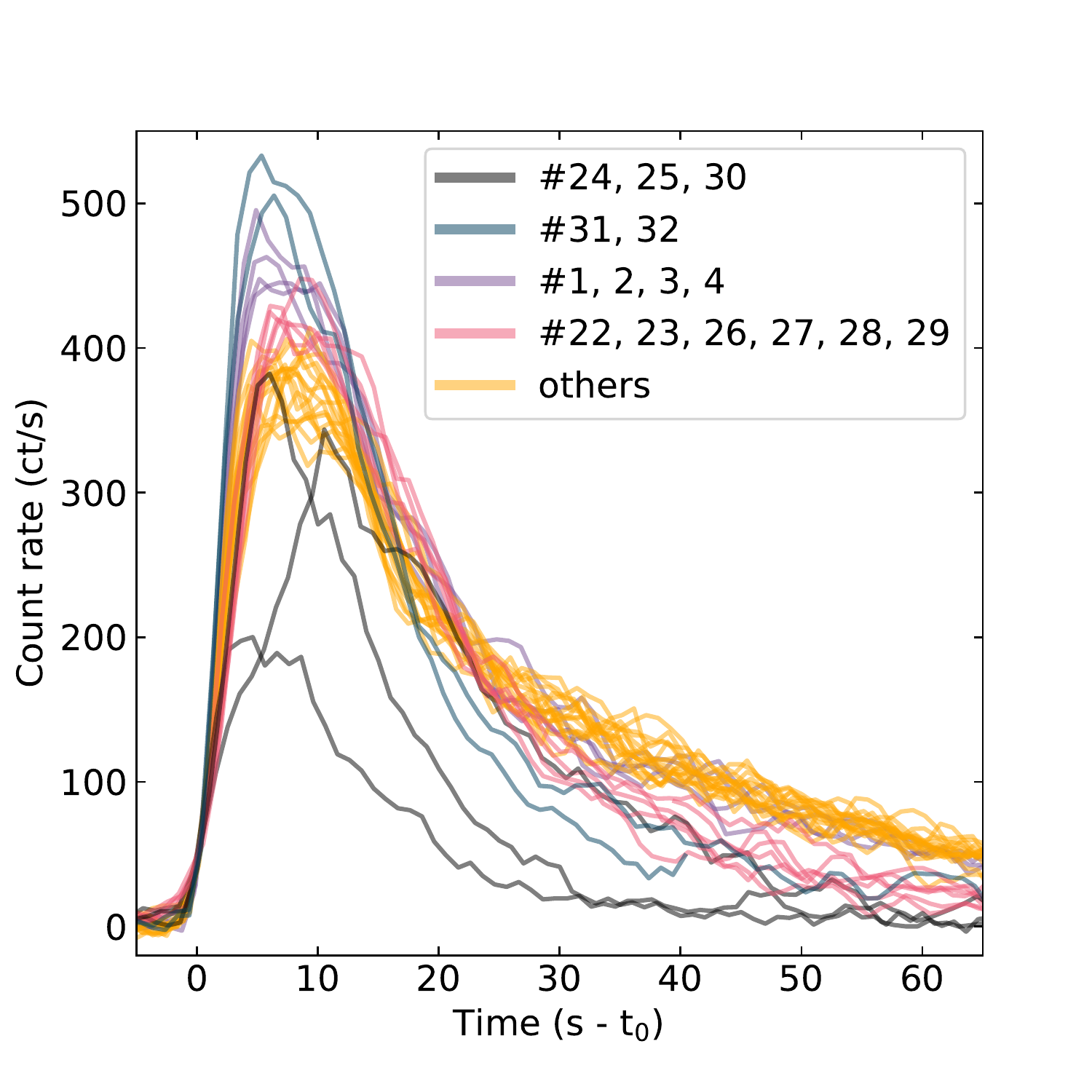}
  \caption{%
	X-ray burst light curves aligned by their start time and with the average
	pre-burst count-rate subtracted. The X-ray burst profiles show a subtle
	evolution: the early and late bursts (lower averaged preburst count rate) reach a 
	higher peak rate and show a faster decay as compared to the other X-ray
	bursts (higher averaged preburst count rates).
  }
  \label{fig:grouped bursts}
\end{figure}

  The similarity between the majority of the X-ray bursts is
  highlighted in Figure \ref{fig:grouped bursts}, where we show all light
  curves in the same graph. 
  We see that the bulk of the burst profiles (yellow curves) are highly
  similar, while the remaining bursts reflect a modest shape evolution across
  the sample.  Specifically, the first four bursts have a sharp rise and higher
  peak count-rates (purple curves). Similarly, bursts \no31 and 32 also peak at
  higher rates than average, and further show a notably shorter decay (blue
  curves). Bursts \no22, 23, and 26-29 (pink curves) show the same tendency as
  \no31-32, but with less pronounced shifts.
  Three X-ray bursts are found to show a deviating profile, and are indicated
  in grey.  Burst \no24 rises much more slowly to its peak intensity, burst
  \no25 peaks at a much lower rate, and burst \no30 decays more rapidly than
  any of the other bursts in the sample.

\subsection{Spectroscopy}
  \label{sec:spectroscopy}

  We study the spectral properties of the X-ray bursts using \textsc{xspec}
  version 12.11 \citep{Arnaud1996} and version 1.02 of the \nicer
  instrumental response matrix. A background spectrum was generated
  using version 0.2 of the ``environmental'' background model\footnote{%
	\url{https://heasarc.gsfc.nasa.gov/docs/nicer/tools/nicer_bkg_est_tools.html}
  } (Gendreau et al., in prep). Interstellar absorption was modelled 
  using the T\"ubingen-Boulder model (\texttt{tbabs}, \citealt{Wilms2000}).

  We first consider the spectral properties of the persistent emission. For
  each X-ray burst we extracted a spectrum from a {$200\s$} window prior
  to the burst onset, keeping at least 50 seconds between the end time of
  the window and the burst onset. In the cases where the burst onset occurred too close to
  the start of the observation, we instead extracted a persistent emission
  spectrum after the burst had decayed instead, selecting the 200-s
  window as close to the end of the observation as possible.
  In all cases the individual $0.5-10\kev$ persistent emission spectra could be
  well described as an absorbed power law.  Applying a joint fit to all
  preburst spectra, we tied the absorption column density across all spectra,
  but let the power law photon index vary per spectrum. We found the
  absorption column density at {$N_\text{H} = \rbr{1.73 \pm 0.01}
  \E{22}\persq{cm}$}, and a power law photon index that gradually increases
  with time from $1.5$ to $2$. We measured $0.5-10\kev$ X-ray flux
  for each spectrum using the \texttt{cflux} model component, and estimated the
  bolometric flux by integrating this component between $0.01-100\kev$. The
  full set of fit parameters and fluxes are listed in the appendix (Table
  \ref{tab:preburst spectra}).

  We analyzed the burst spectra using a time-resolved approach. We adaptively
  binned the burst light curves into multiples of 1 second, such that each bin
  contained at least $500$ counts, yielding about 12 temporal bins per burst. 
  For each bin, we extracted a spectrum and modeled it using an absorbed
  blackbody superimposed on the persistent emission spectrum, holding all
  parameters of the persistent emission spectral model fixed at the values
  listed in Table \ref{tab:preburst spectra}. We also attempted to fit the
  burst spectra while leaving the normalization of the persistent emission as a
  free parameter \citep{Worpel2013}; however, this did not improve the fit, and
  was therefore not pursued further.  We further note that none of the X-ray
  bursts showed evidence for photospheric-radius expansion. 

  After determining the time-resolved spectral parameters, we fit all burst
  spectra again to measure the bolometric flux of the burst emission. We
  multiplied the blackbody component with the \texttt{cflux} model and extrapolated
  the energy range between $0.01-100\kev$. Subsequently, we extracted the highest
  single bin flux and integrated over all time bins in a light curve to measure
  the bolometric fluence of each burst.  The resulting bolometric fluence and
  peak flux measurements of each burst are listed in Table \ref{tab:bursts}.

\begin{figure}[t]
	\centering
	\includegraphics[width=\linewidth]{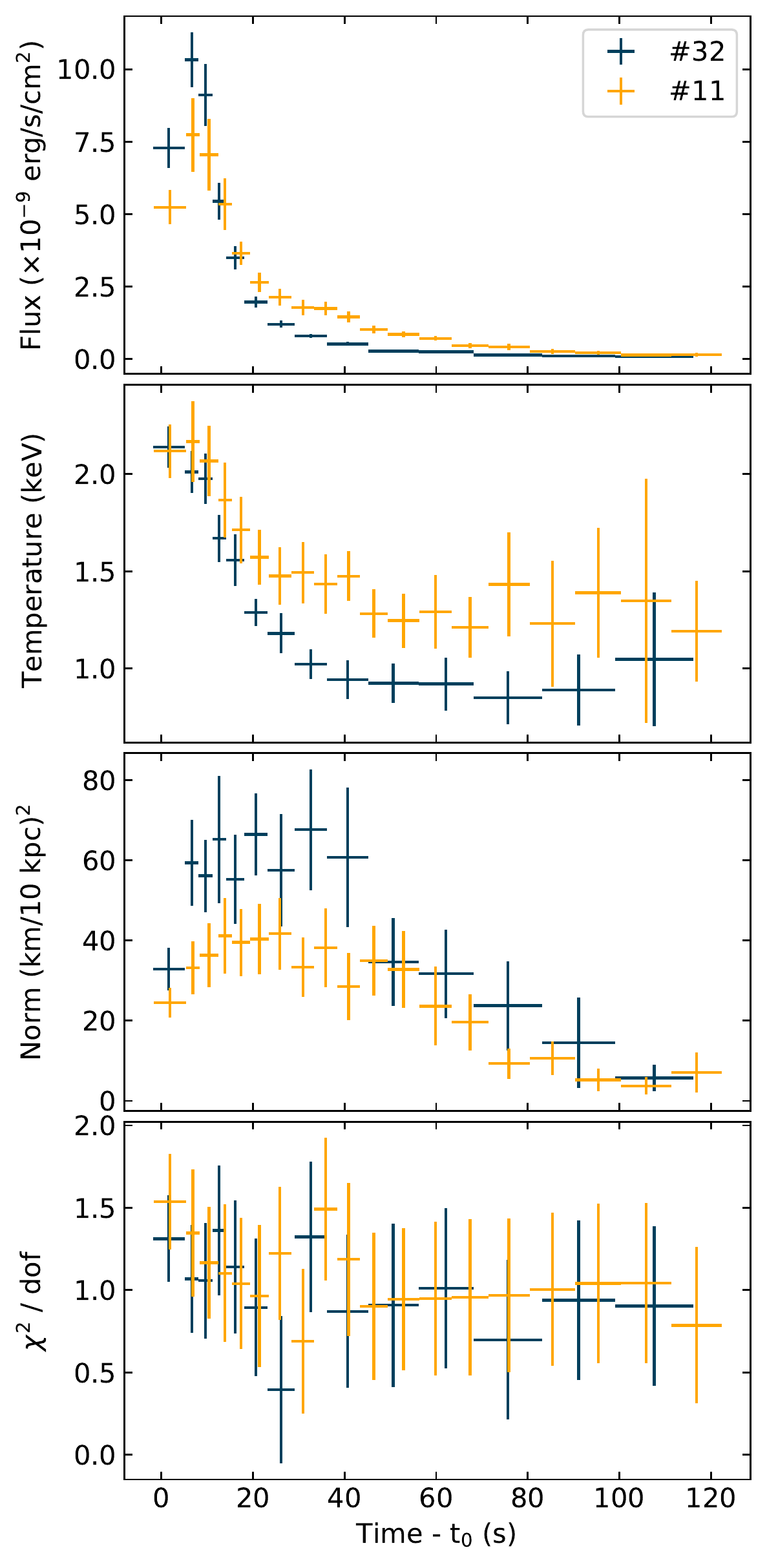}
	\caption{%
		Time resolved spectral fit parameters for X-ray bursts \no11 and \no32, showing
		from top to bottom: the bolometric X-ray flux, the blackbody
		temperature, the blackbody normalization area, and the goodness-of-fit
		statistic. All vertical error bars show the 90\% confidence region,
		which, for the bottom panel, was derived from the $\chi^2$ distribution
		as $1.645 \times \sqrt{2/\text{dof}}$.
	}
	\label{fig:single time resolved}
\end{figure}

\begin{figure}[t]
	\centering
	\includegraphics[width=\linewidth]{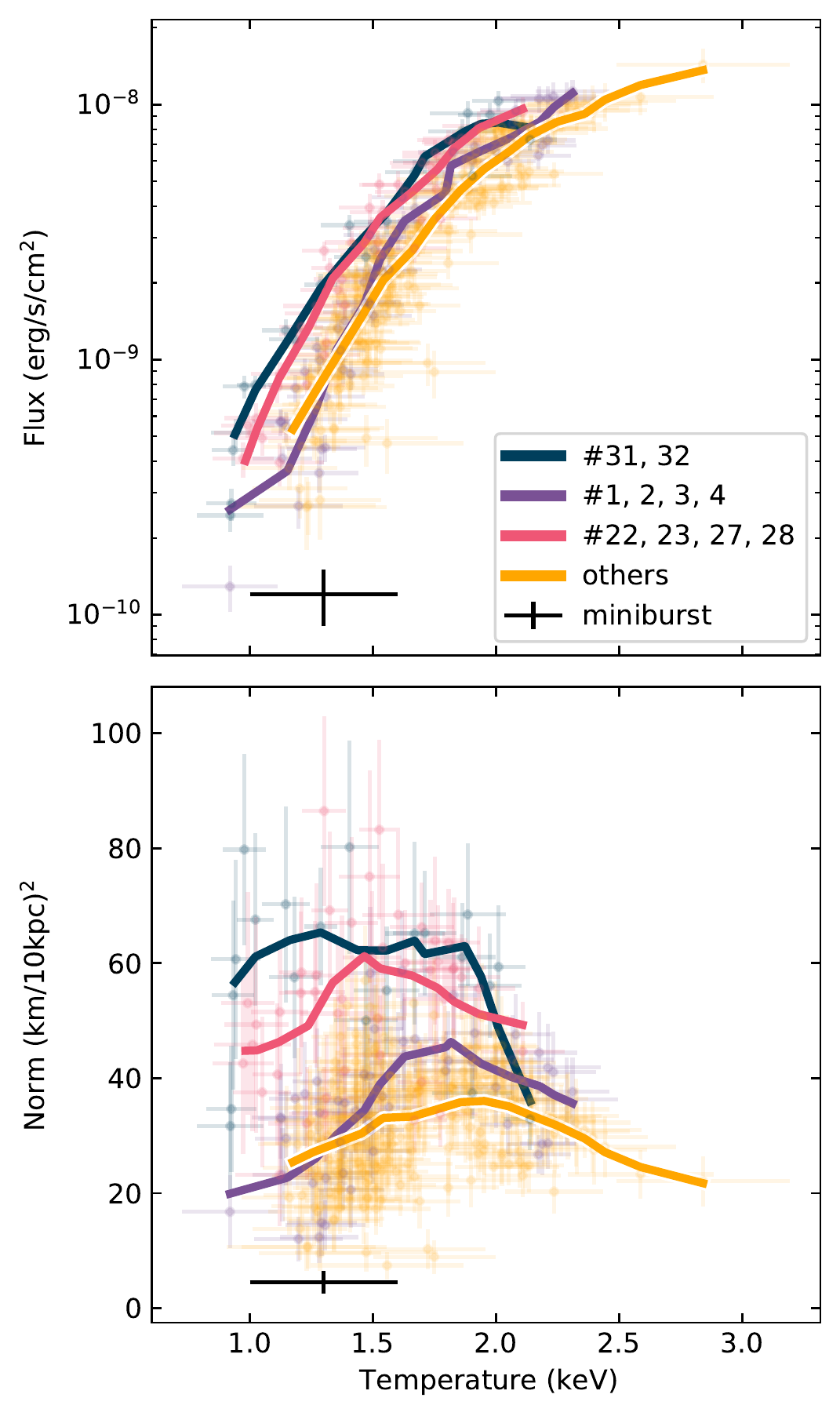}
	\caption{%
		Time resolved spectral fit parameters of all X-ray bursts, showing the
		bolometric X-ray flux (top), and blackbody normalization (bottom) as
		a function of blackbody temperature. The solid lines represent
		the averages for each group. The black point shows the
		mini-burst, otherwise the color coding is the same as in Figure
		\ref{fig:grouped bursts}.
	}
	\label{fig:time resolved}
\end{figure}

  In Figure \ref{fig:single time resolved} we show the evolution of the
  spectral parameters for a typical burst from the largest group (\no11)
  along with the evolution measured for burst \no32. In each case we see that
  the burst emission peaks at a blackbody temperature of about $2.5\kev$, with
  a bolometric flux of $\approx1\E{-8}\fluxcgs$. The blackbody emission area
  peaks about $5-10\s$ later at $40\,({\rm km / 10\,kpc})^2$ and
  $60\,({\rm km / 10\,kpc})^2$ for bursts \no11 and \no32, respectively.
  This evolution pattern is repeated in all X-ray bursts, as illustrated
  in Figure \ref{fig:time resolved}. In this figure we plot the bolometric
  burst flux and normalization area as a function of blackbody temperature
  for all X-ray bursts, showing the average per burst-group with a solid line. 
  Here, we see the lag between respective peaks in blackbody temperature and
  normalization reflected in the curved track traced out in the bottom panel.
  We further see that there is a gradual evolution in the tracks traced out by
  these X-ray bursts. Some of this evolution is even more apparent when
  we consider the peak bolometric flux and peak blackbody temperature as a
  function of time (see Figure \ref{fig:light curve}). When the source
  count-rate climbs above $\approx60\cts$ (bolometric flux
  $\geq1.4\E{-10}\fluxcgs$), both the peak burst flux and temperature decrease
  (bursts \no22-30). 

  Finally, we summarize each burst through three standard X-ray burst
  metrics, as presented in Table \ref{tab:bursts}.
  We measure the e-folding timescale ($\epsilon$) of the burst tails by
  modeling the measured bolometric burst flux evolution as an exponential
  decay.
  We further determine the $\alpha$ factor, which is the ratio of the
  persistent fluence between bursts to the fluence of the burst itself
  \begin{equation}
	\label{eq:measured alpha}
	\alpha = \frac{F_\text{preburst} \Delta t}{E_\text{burst}},
  \end{equation}
  where we estimate the persistent fluence by multiplying the preburst
  flux, $F_\text{preburst}$, with the burst recurrence time, $\Delta t$,
  and $E_\text{burst}$ gives the X-ray burst fluence. 
  Finally, we calculate the ratio of the burst fluence to peak burst flux
  ($\tau$), which represents the equivalent duration of the X-ray burst and
  gives a rough measure of the burst morphology \citep{Paradijs1988}.  We
  calculate these metrics only if the measured recurrence time was less than
  one day, and if the X-ray burst was not truncated. 

\subsection{Mini-burst}
\label{sec:recurrence}
\begin{figure}[t]
	\centering
	\includegraphics[width=\linewidth]{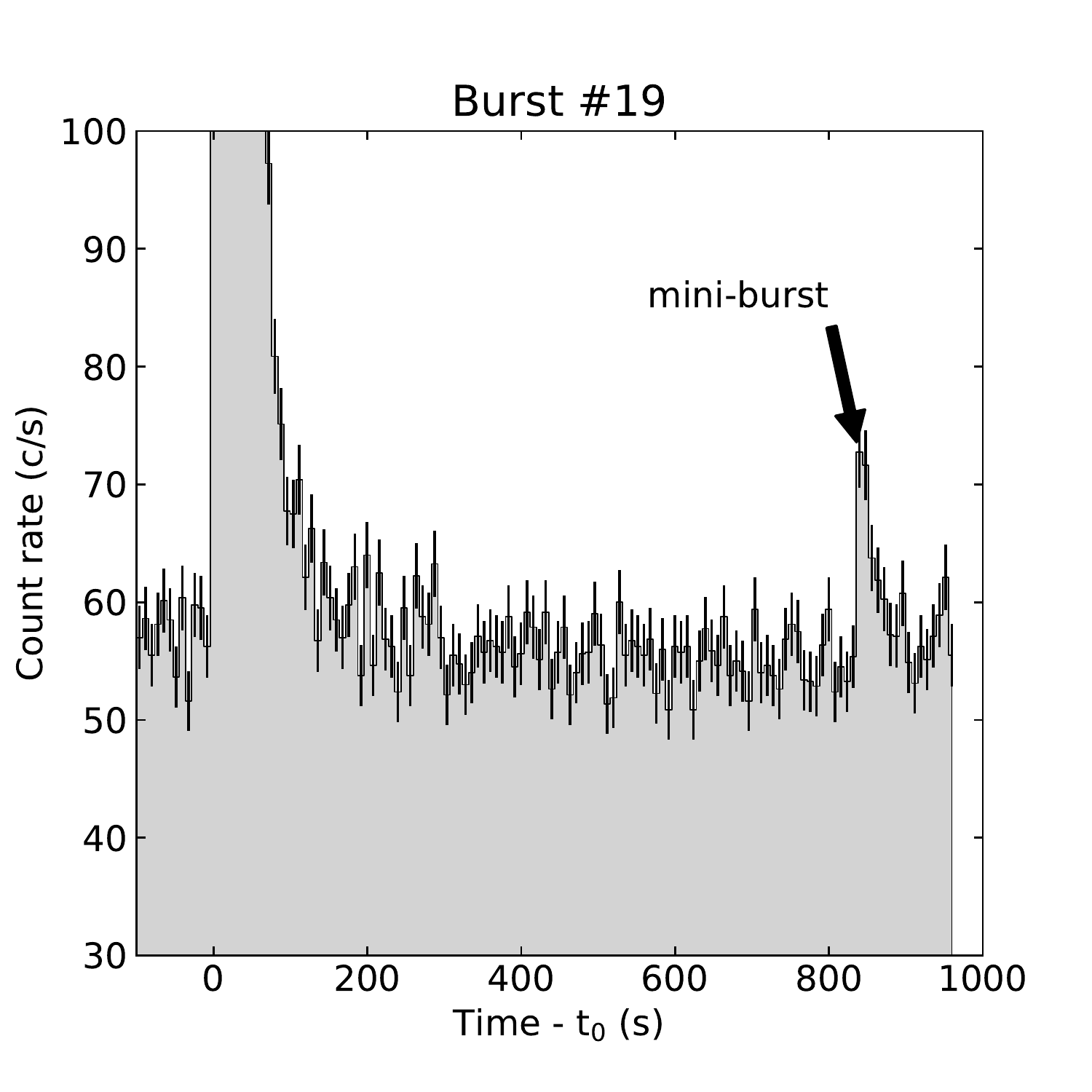}
	\caption{%
		Light curve of the X-ray burst \no19 at 8-s time resolution. A mini-burst
		is observed at $t=840\s$. Errorbars indicate the $1\sigma$ uncertainty. 
	}
	\label{fig:miniburst}
\end{figure}

  In ObsID 3050280106 we observed what we will call a mini-burst. At 840\s after
  the onset of X-ray burst \no19, we observed a brief flare-up in count-rate. The 
  event follows the expected profile of an X-ray burst: it has a sharp rise 
  and an approximately exponential decay. In contrast to regular X-ray bursts, however,
  this mini-burst peaked at 75\cts and lasted only {$\approx30\s$} (see Figure
  \ref{fig:miniburst}). 
  
  We extracted a spectrum for the time interval of the mini-burst and compared
  it with preburst spectrum \no19. Because the mini-burst is comparatively faint,
  the resulting spectrum is of poor quality and can be successfully fit with several
  spectral models. The simplest of such models invokes the power law component
  used to describe the preburst emission, and fits for normalization only.
  However, the burst-like profile of this event invites an interpretation that
  is analogous to the spectral model used for the burst emission. If we fix the
  spectral parameter of the preburst model, then we find that the excess
  emission is well described by a single-temperature blackbody.  The best-fit
  $\chi^2$ is 63 for 61 degrees of freedom, yielding a blackbody temperature of
  $1.3\pm0.3\kev$ with a $0.5-10\kev$ X-ray flux of
  $(1.2\pm0.3)\E{-10}\fluxcgs$. We show this measurement in Figure \ref{fig:time
  resolved} using a black point, and note that this excess emission is
  consistent with the flux-temperature relation measured in Section
  \ref{sec:spectroscopy}.

\subsection{Burst oscillations}
\label{sec:timing}
  We searched the X-ray bursts for the presence of coherent burst oscillations,
  excluding the mini-burst and burst \no29 from the analysis. 
  To avoid confirmation bias towards the 1122\hz burst
  oscillation candidate reported by \citet{Kaaret2007}, we treated our analysis
  as a blind search for an unknown oscillation frequency. Hence, we defined our
  frequency search window to be bounded between $50\hz$ and $2000\hz$. The
  lower bound is motivated by the fact that the non-stationary X-ray
  burst light curve introduces red noise into the power spectrum. For
  frequencies greater than $50\hz$, however, the red noise contribution becomes
  negligible, and the individual frequency bins are well described by a
  $\chi^2$ distribution \citep[see, e.g.][]{Ootes2017, Bilous2019}. The upper
  bound on the frequency range is motivated by physical limits on the maximum
  spin frequency a neutron star can sustain given realistic equation of state
  constraints \citep{Haensel2009}.

  To search for coherent burst oscillations in any given burst, we construct a
  dynamic power spectrum of the $0.5-10\kev$ burst light curve using a sliding
  window method. We apply a window of duration $T$ to a $1/8192\s$ resolution
  light curve and then move this window from $t_0-10\s$ to $t_0+100\s$ in steps
  of $T/2$, where we recall that $t_0$ is the burst onset time (as defined in
  Table \ref{tab:bursts}). For each window position we use the Fourier
  transform to compute the power density spectrum, and extract the highest
  measured power. 
  To establish if this measured power is in excess of the noise, we 
  compare it to two approximations of the noise distribution: the $\chi^2$
  distribution as expected from a pure Poisson counting process; and a numeric
  simulation of the non-stationary burst light curve.

\subsubsection{$\chi^2$ statistics}
  The power spectrum of counting noise is well known to be $\chi^2$ distributed
  \citep{Klis1989}.  For any single frequency bin in the power spectrum, we can
  therefore directly calculate the expected probability that the noise process
  would yield a power greater than the one measured. 
  For every searched X-ray burst, however, we evaluate the powers in
  $N_\text{f}$ frequency bins for $N_\text{w}$ window positions. For simplicity, we treat
  these $N = N_\text{f} \times N_\text{w}$ trials as though they are independent. 
  We can then express the chance that the single-trial survival probability
  $\varepsilon_1$ was produced by noise as $\varepsilon_N = 1 - (1 -
  \varepsilon_1)^N$ \citep[see, e.g.,][]{Vaughan1994}. 
  When the multi-trial survival probability, $\varepsilon_N$, is smaller
  than $1\%$, we consider the measurement to be a detection at the search
  level. Finally, we account for the fact that we search 31 different X-ray bursts
  using multiple search configurations (e.g., different window sizes) by
  increasing the trial count accordingly. 

\subsubsection{X-ray burst simulations}
  A limitation of the $\chi^2$ statistics derived from the counting noise process
  is that it does not account for the fact that the underlying X-ray burst light
  curve is non-stationary, nor for the fact the overlapping window positions impose
  correlations between the measured powers. In an effort to more faithfully account
  for such effects, we also estimate the noise distribution through a series of numeric
  simulations. 

  For a given burst, we generate a sample of artificial light curves using
  an approach similar to that of \citet{Bilous2019}. That is, we wish to generate
  a list of photon arrival times that follows the slow ($\ll50\hz$) variations
  in count-rate of an observed burst, but does not contain any high frequency
  periodic signals. To achieve this, we use the thinning method \citep{Lewis1979}
  to generate a realization of a non-homogeneous Poisson process whose
  underlying rate is specified by a time-continuous light curve. This
  time-continuous light curve, in turn, is constructed from the real data,
  through a linear interpolation on the $1/4\s$ light curve of an observed
  X-ray burst. In short, this procedure involves four steps: 
  \begin{enumerate}
	\item Given the peak count rate, $\lambda_\text{max}=600\cts$, and burst
	  duration, $T=110\s$, draw a random number from the Poisson distribution
	  with mean $T\lambda_\text{max}$, call this random number $N_\lambda$;
	\item Draw $N_\lambda$ arrival times from a uniform distribution on the burst 
	  good time interval $(t_0-10, t_0+100)$, call them $t_i$.
	\item Draw $N_\lambda$ acceptance/rejection criteria from a uniform distribution
	  on $[0,\lambda_\text{max}]$, call them $s_i$.
	\item Keep only those arrival times with $s_i < f(t_i)$, where $f(\cdot)$ is the
	  continuous time light curve. 
  \end{enumerate}

  For each observed X-ray burst, we simulate a set of 5000 analogous
  realizations.  When applying a search procedure to the real data, we
  similarly search the simulated data using the same search method, and extract
  a sample of 5000 `highest' simulated powers. Hence, we directly map out the
  search-level probability distribution (i.e. adjusted for the $N$ trials in
  the dynamic power spectrum).
  Finally, we parameterize each of the resulting distributions into a functional
  form by fitting it with a log-normal distribution. While the choice for
  the log-normal is ad hoc, a Kolmogorov-Smirnov test indicates that this model
  yields a good description (p-value $> 0.05$) of the simulated data in every
  search configuration considered in this paper.

\subsubsection{Sliding window searches}
  We performed a series of sliding window searches on the individual bursts.
  We selected window durations of $2$, $4$, and $8$ seconds and applied them to
  the $0.5-10\kev$ light curves of the bursts. None of these searches yielded a
  power measurement with a chance probability smaller than the $1\%$ noise probability 
  threshold at the search-level. Hence, no significant burst oscillations were
  detected.

  In a second iteration of searches, we used the same sliding window
  configuration, and additionally applied a factor 4 binning to the power
  spectra. This approach is motivated by the fact that burst oscillations may
  drift in frequency over the course of an X-ray burst, in which case the signal
  power may be spread out across several frequency bins \citep[see, e.g.,][for
  a review]{Watts2012}. 
  
\begin{figure}[t]
	\centering
	\includegraphics[width=\linewidth]{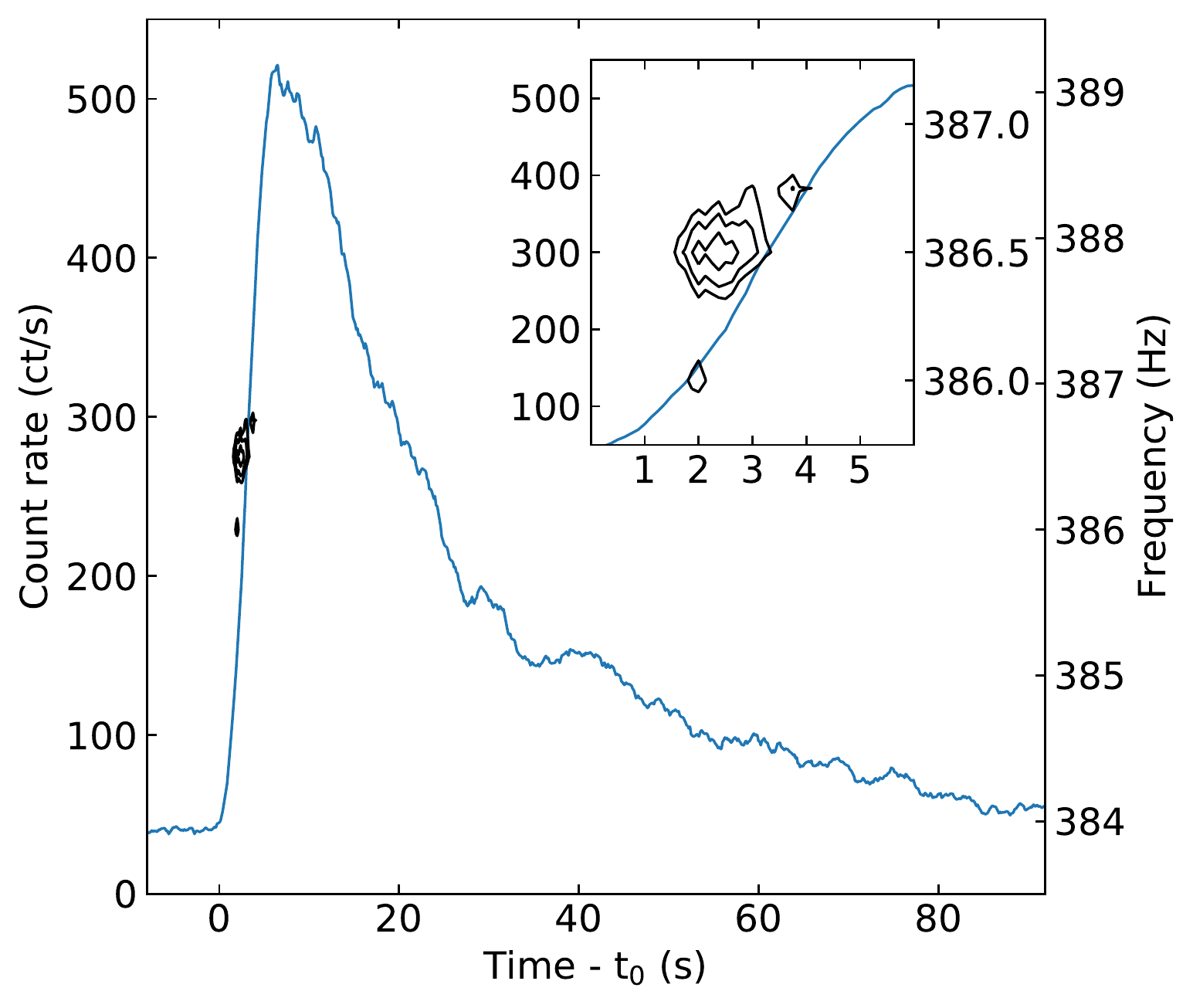}
	\caption{%
		Burst oscillation detection in X-ray burst \no2, showing the contours
		of a dynamic power spectrum (black, right axis), along with the burst
		light curve (blue, left axis). The contours mark the power
		(17.4, 21.1, and 26.8) corresponding to the 68\%, 95\% and 99.7\% 
		confidence levels, adjusted for the number of trial frequencies in the spectrum. 
		For illustrative purposes, we oversampled the dynamic power spectrum,
		i.e., it was calculated using 4\s time windows with steps of 1/8\s. The
		inset shows the same data, but zoomed-in on the region where the burst oscillation candidate was detected.
	}
	\label{fig:burst oscillation}
\end{figure}

  Applying the searches with binning in the frequency domain, we recovered a
  candidate 386.5\hz burst oscillation in X-ray burst \no2.  The candidate
  signal occurred during the rise of the burst (see Figure \ref{fig:burst oscillation})
  and is observed for all three window durations. Based on $\chi^2$ statistics,
  the highest measured power corresponds to a single-trial noise probability of
  $1.2\E{-12}$.  Accounting for the number of trials in all six search
  configurations (three with, and three without frequency binning), we obtain
  a multi-trial adjusted noise probability of $1.8\E{-6}$. Further adjusting
  the trial count to include all 31 X-ray bursts, we obtain $5.6\E{-5}$. Hence,
  the chance probability that this signal is produced by the counting noise is
  well below the $1\%$ threshold. Comparing the signal with the numerically
  estimated noise distribution for X-ray burst \no2, we find that the six
  search configurations have a joint probability of $9\E{-5}$ to produce the
  observed power by chance. Again extending this analysis to include all
  31 X-ray bursts, the total joint noise probability is $2.8\E{-3}$. Hence, the 
  386.5\hz signal is again found to be in excess of the noise process, albeit
  at a lower level of significance.

  Given the frequency of the detected burst oscillation in burst \no2, we can return
  to the other bursts to search with higher sensitivity. We repeated all six search
  procedures, searching only the narrow frequency range from 377\hz to 397\hz.
  This third iteration of searches did not yield any additional detections. 

  The Leahy normalized power spectrum of a Poisson sampled coherent wave yields
  a power distribution that is well described as a non-central $\chi^2$
  distribution (see, e.g.,  \citealt{Groth1975}). We write this function as
  $\overline\chi^2(P_\text{m} | \kappa, \zeta)$, with $P_\text{m}$ the measured
  power, $\kappa$ the degrees of freedom, and $\zeta$ the non-centrality
  parameter. The latter depends on the signal amplitude, $A$, as 
  \begin{equation}
  	\zeta=\frac{\kappa A^2} {4N_\gamma},
  \end{equation}
  with $N_{\gamma}$ the number of photons. We numerically invert this
  relation to find the distribution of the signal amplitude given the measured
  power, and finally express the burst oscillation amplitude as a fractional
  sinusoidal amplitude relative to the burst flux
  \begin{equation}
	a = \frac{A}{N_\gamma - N_\text{persistent}},
  \end{equation}
  where $N_\text{persistent}$ gives the number of photons contributed by the
  persistent emission, as estimated from the preburst count-rate. Using this
  formalism, we find a fractional amplitude of $a = (26\pm4)\%$ for the candidate
  oscillation in burst \no2.

\section{Discussion}
\label{sec:discussion}
We have presented a spectral and timing analysis of 32 Type I X-ray bursts
observed from \src with \nicer. All X-ray bursts were detected over a 20-day
period during which the persistent X-ray emission increased from
$4.5\E{-10}\fluxcgs$ to $1\E{-9}\fluxcgs$. This flux range is broadly
consistent with the source intensity at which X-ray bursts have been previously
reported \citep[Section \ref{sec:minbar}]{ATelBrandt05,Kaaret2007}, and
is associated with the hard state of this source (see Section
\ref{sec:state}).

\subsection{X-ray burst energetics}
The profiles of the X-ray burst light curves are mostly very similar: they take
$5-7$\s to rise to their peak intensity, and are followed by an
approximately 100\s decay. Such profiles indicate the bursts are due to helium
burning in a hydrogen rich environment \citep[see, e.g.,][and references
therein]{Galloway2017b}, with a cooling tail that is governed by the rp-process
\citep{Schatz2001}.

Another view of the burst fuel composition is given by the $\alpha$
factor. Due to the relatively short exposures of individual \nicer pointings,
we did not observe successive bursts in a single uninterrupted exposure,
such that all burst recurrence times and $\alpha$ factors listed in Table
\ref{tab:bursts} are formally upper limits. Instead, we estimate the
averaged bursting rate by dividing the 226\ks unfiltered exposure collected
between the first and last ObsIDs containing an X-ray burst by the number of
bursts observed, yielding a recurrence time of $2.0_{-0.3}^{+0.4}\hr$. 
If we adopt average values for the burst fluence, persistent flux, and burst
waiting time, we find an average $\alpha$ of 47, which is consistent
with a mixed hydrogen/helium fuel composition.

If we assume that each X-ray burst depletes the available reservoir of fuel,
then $\alpha$ can be predicted from theory as
\begin{equation}
  \alpha = \frac{Q_\text{grav}}{Q_\text{nuc}} (1+z) 
	  \frac{\xi_\text{b}}{\xi_\text{p}},
\end{equation}
where $Q_\text{grav} = GM_\text{NS} / R_\text{NS}$ is gravitational potential
energy released through accretion, with $G$ the gravitational constant,
$M_\text{NS}=1.4\msol$ the neutron star mass, and
$R_\text{NS}=10\km$ the neutron star radius. Additionally,
$Q_\text{nuc} = 1.35 + 6.05\overline{X}\,\mbox{MeV/nucleon}$
\citep{Goodwin2019a} is the nuclear energy generation rate in a burning layer
with averaged hydrogen fraction, $\overline{X}$, and $(1+z)=1.31$ gives
the gravitational redshift factor. Finally, $\xi_\text{p}$ and $\xi_\text{b}$
give the anisotropy factors for the persistent and burst emission,
respectively. Given the observed $\alpha$ and an allowed range for hydrogen
abundance of $0.1
< \overline X < 0.7$, we find that the ratio of anisotropy factors,
$\xi_\text{b}/\xi_\text{p}$, is in the range $0.3-0.9$. 
If we adopt a simple thin disk model, then we can relate these anisotropy
factors to the system inclination \citep{Fujimoto1988, He2016}, such that the
allowed system inclination is $65\arcdeg < i < 90\arcdeg$. Hence, the bursting
properties suggest that \src is a relatively high inclination system. The lack
of eclipses in the light curve further indicates that we are not viewing the
system edge-on, allowing for an upper limit on the inclination of $i \lesssim
75\arcdeg$ \citep{Frank1987}.

\subsection{Burst light curve variations}

While the majority of observed bursts are very similar, some evolution
in the burst light curve shapes can be observed across the sample. This is
demonstrated in Figure \ref{fig:grouped bursts}, where we grouped the burst light
curves by shape. Relative to the most commonly observed burst shape (yellow),
we selected three groups of burst shapes that each tend toward higher peak count-rates
and shorter decay tails (blue, purple, pink). 
In addition to exhibiting similar burst profiles, these bursts share commonalities
in time and persistent flux as well. The bursts of the first group (\no1-4; purple curves) occur early in
the outburst and have the lowest bolometric persistent fluxes of the sample ($8\E{-10}\fluxcgs$).
Those in the second group (\no22-29, pink curves) have the highest persistent fluxes
in the sample ($1.5-1.8\E{-9}\fluxcgs$), and coincide with a notable drop in peak
burst temperature and peak burst flux (Figure \ref{fig:light curve}). Finally,
the third group bursts (\no31, 32) have the sharpest profiles of the observed bursts, and occur
right after the X-ray flux dropped back down to $1\E{-9}\fluxcgs$.

The observed pivot in burst shape from longer burst with lower peak
rates to shorter bursts with higher peak rates is not uncommon and can be
attributed to change in fuel composition, with sharper bursts having a
comparatively higher helium abundance \citep{Lewin1993, Galloway2017b}.
Some evidence that this effect is at play in \src can be found in the $\alpha$ factors. 
Considering the $\alpha$ factors of those bursts with the most robust
recurrence times, we find values ranging from about 40 (\no12,14) to
80 (\no29) and 134 (\no30), suggesting that the bursts with sharper
profiles indeed have a lower hydrogen content.

What physical process is driving fuel composition changes in \src is
less clear. The most likely driver of such evolution is the changing mass accretion rate
\citep[e.g.,][]{Bildsten1998a}.
However, the most common burst shape has the longest tails, and thus the highest
hydrogen content. Relative to this group, we see that the hydrogen content
decreases for both increasing and decreasing mass accretion rates.

\subsection{Short-recurrence bursts} \label{sec:swt}
A caveat to the interpretation of $\alpha$ is that at least some of the X-ray
bursts do not appear to exhaust the fuel layer. Given its light curve profile
and short waiting time relative to burst \no19, the mini-burst reported in
section \ref{sec:recurrence} is most likely a short-recurrence burst
\citep[see, e.g.,][]{Keek2010}. Such bursts occur with recurrence times that
are too short for the accretion process to fully replenish the stellar
atmosphere, and are therefore fuelled by left-over hydrogen and helium on the
stellar surface. 

If the mini-burst is indeed a short-recurrence burst, then it seems likely that
X-ray burst \no25 was also a short recurrence event. This burst occurred 480\s
into its respective pointing, and is preceded by a 6\,hr gap in coverage. Given
that the burst recurrence time is on the order of 2 hours, we almost assuredly
missed the X-ray bursts directly preceding burst \no25.

In a similar vein, we can also consider bursts \no24 and \no30 in the
context of short recurrence events. Each of these two bursts were 
less energetic than the remaining sample, albeit not to the extent of burst \no25.
If these bursts only burned some fraction of the available fuel, that would account
for their reduced intensity, while leaving requisite material behind for short recurrence
bursts \citep{Keek2010, Keek2017b}. Hence, we can speculate that bursts \no24
and \no30 were primary events of short recurrence trains. These bursts were followed
by only 7 and 9 minutes of exposure before a 66 and 78 minute data gap, respectively.
The data sampling therefore leaves sufficient space for such short recurrence events
to have occurred. This interpretation is complicated by the slow rise of burst \no24,
as the process that governed this slow rise might also underpin the reduced intensity.
It is not clear what causes this slow rise, although we note that ignition
latitude has been linked to the rise morphology \citep{Maurer2008}.

\subsection{The MINBAR sample}
\label{sec:minbar}

\begin{figure*}[t]
	\centering
	\includegraphics[width=\linewidth]{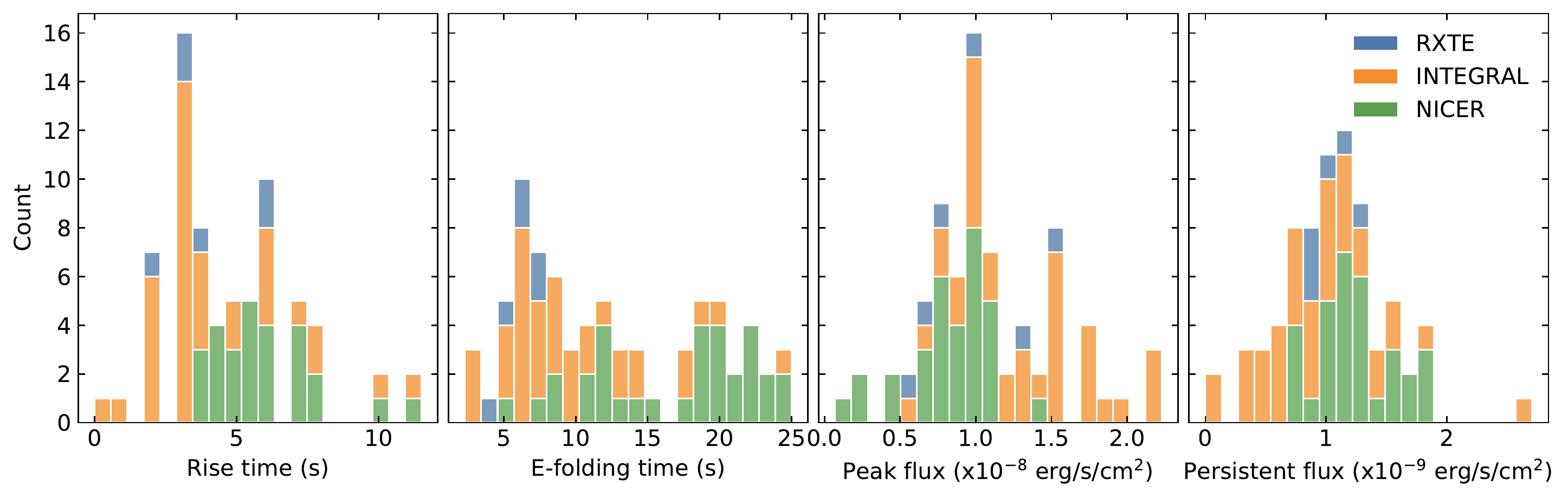}
	\caption{%
		Stacked histograms of X-ray burst properties measured with \nicer versus those reported in MINBAR.
	}
	\label{fig:minbar}
\end{figure*}

Most of the X-ray bursts observed with \nicer appear to have an
appreciable hydrogen content. This stands in contrast with the X-ray bursts
observed with \rxte, which were all reported to be typical helium-fueled bursts
\citep{Kaaret2007, Galloway2008}. This raises the question if the bursting
regime sampled with \nicer is somehow different than the bursting phases
observed previously. To investigate this question, we compare some of the burst
statistics obtained with \nicer to those reported in MINBAR \citep{Galloway2020}. 
In Figure \ref{fig:minbar} we show histograms of the burst rise time,
e-folding time, peak burst flux and the persistent flux. Each is color coded per
observatory (blue for \rxte, orange for \integral, and green for \nicer). We
see that some systematic shifts are apparent: the \nicer bursts appear to show
a slower rise and longer decay than the \rxte and \integral bursts. The \nicer
peak fluxes are slightly lower, while the flux of the persistent accretion is
marginally higher. On the whole, however, the MINBAR and \nicer samples are not
dramatically different.

\subsection{Spectral state}
\label{sec:state}
\begin{figure}[t]
	\centering
	\includegraphics[width=\linewidth]{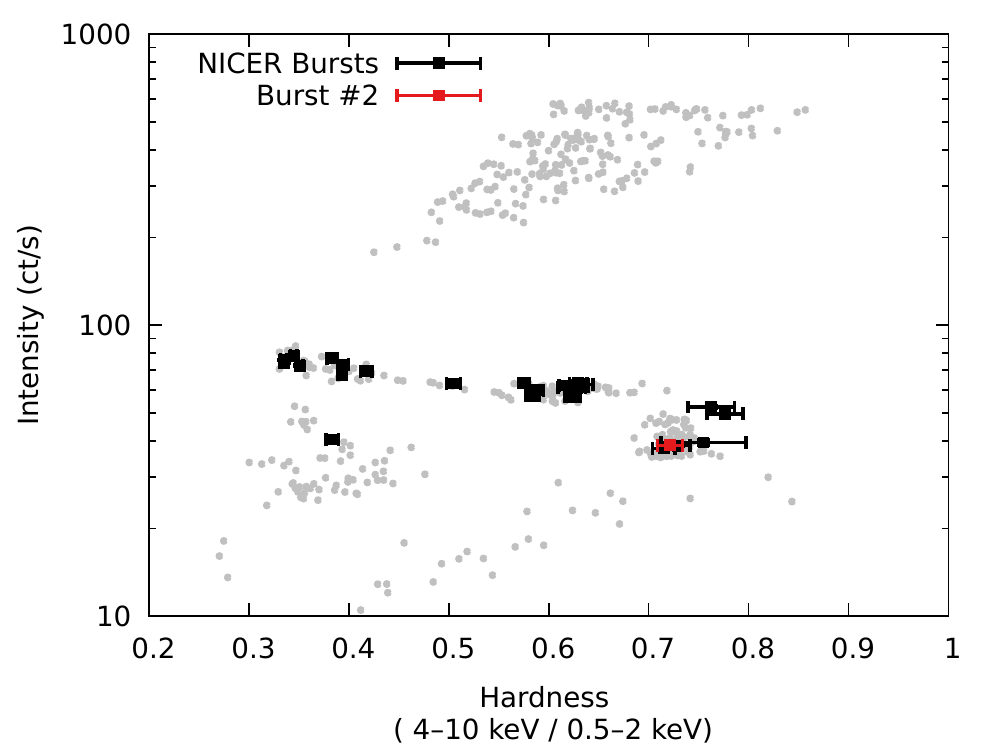}
	\caption{%
		Hardness-intensity diagram for \src, showing one data point for each continuous \nicer pointing in grey. We further highlight those pointings containing an X-ray burst in black, and show burst \no2 in red. For visual clarity we show the $1\sigma$ error bars for the highlighted points only.
	}
	\label{fig:hid}
\end{figure}

To place our observations in the context of the source state and accretion rate
evolution, we constructed a hardness intensity diagram from the light curve
data described in Section \ref{sec:light curves} (Figure \ref{fig:hid}). That
is, we calculated the hardness ratio as the $4-10\kev$ count rate over the
$0.5-2\kev$ count rate and adopted the full band ($0.5-10\kev$) rate for the
intensity, averaging the data per continuous pointing. In addition to the
bursting epoch observations, we also included the \nicer data collected in
August 2019 \citep{ATelBult19e}, which covered higher intensities (count-rates
$>100\cts$). The X-ray bursts are clearly observed along a hard state track.
Toward the highest persistent accretion rates of the bursting epoch, however,
the source may have transitioned to an intermediate state.

Considering the time-evolution of the source intensity and hardness, as shown
in Figure \ref{fig:light curve}, we see that the highest observed count-rates
occur after a rapid drop in hardness. Coincident with this apparent spectral
shift, the X-ray bursts show a decrease in their peak bolometric flux and peak
blackbody temperature. What causes this apparent change in burst character is
not obvious, but it might be explained by a change in the accretion state. 

One possibility is that a change in the accretion disk structure is affecting
the anisotropy ratio. For instance, the formation of a surface boundary layer
could cause more of the burst emission to be shadowed out by accretion flow,
thus lowering the observed burst flux. Such shadowing, however, is predicted to
cause a hardening of the burst spectrum \citep{Suleimanov2012, Kajava2014},
which is opposite to what we observe. 

Alternatively, the decrease in burst flux and temperature might be
related to the presence of short-recurrence bursts.  When
short-recurrence bursts occur, they change the fuel layer, and alter ignition
conditions of long-duration bursts as well \citep{Keek2017b}. Possibly this
effect is what drives the depression in the peak burst flux and temperature.
This scenario is compatible with our data, as all candidate short-recurrence
bursts discussed Section \ref{sec:swt} occur during this short time interval
where the count-rate is high and the hardness is low. Furthermore, surveys of
X-ray bursters show that short-recurrence bursts often only occur in a
comparatively narrow range of luminosity just below the transition to the soft
state \citep{Keek2010}.

To estimate the source mass accretion rate during the bursting epoch, we adopt
a source distance of $4_{-2}^{+4}$\kpc, as derived from Gaia parallax
measurements \citep{Bailer2018}. Assuming a 1.4 solar mass neutron star with
a 10\km radius, the measured bolometric persistent flux implies a mass
accretion rate of $\dot M \approx 2\E{-10}\,\msol\per{yr}$, or about 1.5\% of
the Eddington rate. This accretion rate is lower than the $\approx 10\%$
Eddington rate expected based on the burst behavior \citep{Galloway2008,
Galloway2017b}. While discrepancies between the accretion rate inferred from
the X-ray luminosity and the burst physics are not uncommon \citep[see,
e.g.,][]{Cornelisse2003}, there are additional caveats to this comparison. First,
the Gaia distance is likely an underestimate. As noted by \citet{Galloway2020},
the distance is derived using a probabilistic method that is weighted on the
distribution of matter in the galaxy. Because LMXBs are likely more
concentrated toward the galactic center than the stellar population, however,
the resulting Gaia distance is likely biased. Second, in estimating the
accretion rate, we did not account for the anisotropy associated with the
system inclination. The $\alpha$ factor indicates that we are likely viewing the
binary at very high inclination, such that accretion luminosity is
preferentially beamed away from the line of sight. Thus, the accretion rate
onto the neutron star is likely higher than inferred.

\subsection{Stellar spin frequency}
We searched 31 of the observed X-ray bursts for a presence of burst
oscillations, finding one candidate signal at a frequency of 386.5\hz. With
a single trial probability of $10^{-12}$, the detected signal deviates
significantly from the expected noise distribution. After accounting for all
frequencies tested across all bursts, we conservatively estimated the
multi-trial adjusted probability at $2.8\E{-3}$. 

The signal was detected only after binning the power spectra by a factor of
four, suggesting that the underlying oscillation had a drifting frequency.
Considering the sub-threshold detections around the peak power shows weak
evidence that the signal is drifting by about 1\hz in the $\approx 2\s$ period
that the power is highest. Considering that the candidate detection occurs
during the rising slope of X-ray burst \no2, such a drift in frequency would be
in line with known behaviors of burst oscillations \citep[see,
e.g.,][]{Watts2012}.

The fractional amplitude of the signal is measured at $26\pm4\%$. While such an
amplitude is plausible from theoretical considerations \citep{Mahmoodifar2016},
the observational record indicates that it is comparatively large
\citep{Ootes2017, Galloway2020}. While rising phase burst oscillations have
systematically larger amplitudes than cooling tail oscillations, the amplitude
also depends on the accretion rate. Similarly strong oscillations are normally
observed only in the soft state \citep{Ootes2017, Galloway2020}. In contrast,
the burst oscillation candidate detected in this work occurred in the hard
state. 

In addition to showing different amplitudes, the detectability of burst
oscillations also depends on the source spectral state. Burst oscillations can
occur at any accretion rate, however the fraction of bursts that show such
oscillations has been found to be much higher in the soft state than in the
hard state \citep{Muno2004, Ootes2017, Galloway2020}. This tendency may go some
way to explain why we only detected one candidate signal out of 31 analyzed
bursts.  A caveat, however, is that all known systematics of burst oscillations
are derived entirely from observations made with \rxte, and thus based on an
higher energy passband than that of \nicer. A limited number of burst
oscillation detections made with \nicer \citep{Mahmoodifar2019, Bult2019b} do
not yet point to dramatically different behavior at lower energies. Instead,
because the burst oscillation amplitudes increase with photon energy, it
appears to be more challenging to observe these signals with \nicer than it was
with \rxte. This energy dependence might also play a role in explaining why the
386.5\hz signal was only detected in a single X-ray burst.

While the 386.5\hz burst oscillation candidate deviates significantly from the
noise distribution, it is only observed in a single independent time-frame and
does not repeat in any of the other bursts.  So, while all observed
characteristics make it a plausible burst oscillation, we caution that this
signal should be treated as a candidate until it can be independently
confirmed.  Even with all noted caveats, it is worth stressing that the
386.5\hz signal reported here is more prominent than the 1122\hz oscillation
candidate reported by \citet{Kaaret2007}, which had a single trial probability
of $5\E{-10}$.  The ratio of the two frequencies is 2.9, so it is possible that
the two signals are harmonically related. However, given that such a relation
prefers a fundamental frequency smaller than 386.5\hz and that burst
oscillations tend to appear at or below the underlying spin frequency
\citep{Watts2012}, such an interpretation does not appear probable. Either way,
based on the NICER observations presented here, it seems unlikely that \src has a
sub-millisecond spin period.

    ~\\

%\nolinenumbers

\acknowledgments
This work was supported by NASA through the \nicer mission and the
Astrophysics Explorers Program, and made use of data and software 
provided by the High Energy Astrophysics Science Archive Research Center 
(HEASARC).
D.A. acknowledges support from the Royal Society. 
T.G. has been supported in part by the Scientific and Technological Research Council (T\"UBITAK) 119F082, Royal Society Newton Advanced Fellowship, NAF$\backslash$R2$\backslash$180592, and Turkish Republic, Directorate of Presidential Strategy and Budget project, 2016K121370.
C.M. is supported by an appointment to the NASA Postdoctoral Program at the
Marshall Space Flight Center, administered by Universities Space Research
Association under contract with NASA.

\facilities{ADS, HEASARC, NICER}
\software{heasoft (v6.27.2), nicerdas (v7a)}

\bibliographystyle{fancyapj}

\appendix
\restartappendixnumbering
\section{Spectroscopy}
\begin{table}[h]
  \caption{%
	Preburst spectral parameters
	\label{tab:preburst spectra}
  }
  \centering
  \newcommand{\nh}{\multirow{32}{*}{$1.73\pm0.01$}}
  \hspace*{-1.2cm}
  \begin{tabular}{c c c c c c}
	Burst & $N_\text{H}$        & Photon Index & X-ray Flux			  & Bol. Flux & $\chi^2$/bins \\
	~     & (\E{22}\persq{cm}) & ~            & ($\E{-9}\fluxcgs$) &($\E{-9}\fluxcgs$) \\
	\tableline
	 1 & \nh & $1.49 \pm 0.04$ & $0.451 \pm 0.010$ & $0.77 \pm 0.02$ & 253.96 / 200 \\
	 2 & ~   & $1.50 \pm 0.04$ & $0.464 \pm 0.010$ & $0.79 \pm 0.02$ & 170.85 / 197 \\
	 3 &     & $1.48 \pm 0.04$ & $0.460 \pm 0.010$ & $0.78 \pm 0.02$ & 209.59 / 195 \\
	 4 &     & $1.48 \pm 0.07$ & $0.457 \pm 0.016$ & $0.78 \pm 0.03$ & \phantom{0}94.70 / \phantom{0}79 \\
	 5 &     & $1.45 \pm 0.04$ & $0.585 \pm 0.013$ & $1.00 \pm 0.03$ & 164.74 / 184 \\
	 6 &     & $1.48 \pm 0.04$ & $0.586 \pm 0.011$ & $1.00 \pm 0.02$ & 239.28 / 238 \\
	 7 &     & $1.52 \pm 0.04$ & $0.611 \pm 0.011$ & $1.03 \pm 0.02$ & 254.14 / 254 \\
	 8 &     & $1.45 \pm 0.05$ & $0.631 \pm 0.016$ & $1.08 \pm 0.04$ & 130.10 / 141 \\
	 9 &     & $1.62 \pm 0.06$ & $0.768 \pm 0.022$ & $1.27 \pm 0.04$ & \phantom{0}95.82 / 106 \\
	10 &     & $1.61 \pm 0.03$ & $0.762 \pm 0.012$ & $1.27 \pm 0.02$ & 295.87 / 291 \\
	11 &     & $1.64 \pm 0.03$ & $0.733 \pm 0.012$ & $1.21 \pm 0.02$ & 285.46 / 287 \\
	12 &     & $1.63 \pm 0.03$ & $0.739 \pm 0.012$ & $1.23 \pm 0.02$ & 317.18 / 286 \\
	13 &     & $1.60 \pm 0.03$ & $0.732 \pm 0.012$ & $1.22 \pm 0.02$ & 254.77 / 276 \\
	14 &     & $1.62 \pm 0.03$ & $0.747 \pm 0.012$ & $1.24 \pm 0.02$ & 280.75 / 289 \\
	15 &     & $1.71 \pm 0.03$ & $0.755 \pm 0.012$ & $1.25 \pm 0.02$ & 331.93 / 292 \\
	16 &     & $1.65 \pm 0.03$ & $0.727 \pm 0.012$ & $1.20 \pm 0.02$ & 311.73 / 283 \\
	17 &     & $1.65 \pm 0.03$ & $0.719 \pm 0.012$ & $1.19 \pm 0.02$ & 288.83 / 283 \\
	18 &     & $1.67 \pm 0.03$ & $0.702 \pm 0.012$ & $1.16 \pm 0.02$ & 274.24 / 274 \\
	19 &     & $1.63 \pm 0.04$ & $0.684 \pm 0.013$ & $1.13 \pm 0.02$ & 206.22 / 218 \\
	20 &     & $1.68 \pm 0.04$ & $0.686 \pm 0.011$ & $1.13 \pm 0.02$ & 254.35 / 272 \\
	21 &     & $1.83 \pm 0.04$ & $0.742 \pm 0.012$ & $1.24 \pm 0.02$ & 321.53 / 286 \\
	22 &     & $1.99 \pm 0.03$ & $0.833 \pm 0.013$ & $1.46 \pm 0.03$ & 324.82 / 288 \\
	23 &     & $2.05 \pm 0.03$ & $0.866 \pm 0.014$ & $1.57 \pm 0.04$ & 263.70 / 293 \\
	24 &     & $2.07 \pm 0.05$ & $0.929 \pm 0.020$ & $1.70 \pm 0.06$ & 153.40 / 191 \\
	25 &     & $2.05 \pm 0.04$ & $0.839 \pm 0.014$ & $1.52 \pm 0.04$ & 289.66 / 295 \\
	26 &     & $2.08 \pm 0.03$ & $0.860 \pm 0.014$ & $1.58 \pm 0.05$ & 302.44 / 293 \\
	27 &     & $2.14 \pm 0.03$ & $0.864 \pm 0.015$ & $1.65 \pm 0.05$ & 250.30 / 287 \\
	28 &     & $2.14 \pm 0.03$ & $0.894 \pm 0.016$ & $1.78 \pm 0.06$ & 350.75 / 295 \\
	29 &     & $1.97 \pm 0.02$ & $1.026 \pm 0.012$ & $1.79 \pm 0.03$ & 427.82 / 433 \\
	30 &     & $2.14 \pm 0.03$ & $0.954 \pm 0.016$ & $1.82 \pm 0.06$ & 275.86 / 301 \\
	31 &     & $2.12 \pm 0.04$ & $0.554 \pm 0.011$ & $1.04 \pm 0.04$ & 266.98 / 222 \\
	32 &     & $2.07 \pm 0.04$ & $0.469 \pm 0.010$ & $0.86 \pm 0.03$ & 240.30 / 201 \\
	\tableline
  \end{tabular}
  \flushleft
  \tablecomments{%
	Reported X-ray fluxes are unabsorbed and measured in the $0.5-10\kev$ band.
	Bolometric fluxes are estimated by
	extrapolating the spectral model between $0.01-100\kev$. All uncertainties are quoted at
	90\% confidence. The right-most column of the table lists the $\chi^2$ and number
	of bins contributed by each spectrum. Adding this column together, we obtain a
	joined $\chi^2$ fit statistic of {$8182.10$} for {$7965$} degrees of
	freedom.
  }
\end{table}

\end{document}